\documentclass[conference,9pt]{IEEEtran}
\IEEEoverridecommandlockouts
\usepackage{cite}
\usepackage{amsmath,amssymb,amsfonts}
\usepackage{graphicx}
\usepackage{textcomp}
\usepackage{amsmath}
\usepackage{amsmath}
\usepackage[margin=1in]{geometry}

\usepackage[linesnumbered,ruled]{algorithm2e}% 
\usepackage{algpseudocode}
\usepackage{amsmath}
\usepackage{amsfonts}
\usepackage[margin=1in]{geometry}
\usepackage{float}

\usepackage{caption}
\usepackage{subcaption}
\usepackage{ragged2e} % for '\Centering' macro
\usepackage{booktabs} 
\usepackage{multirow}
\usepackage{hyperref}       % hyperlinks
\usepackage{url}            % simple URL typesetting
\usepackage{booktabs}       % professional-quality tables
\usepackage{amsfonts}       % blackboard math symbols
\usepackage{booktabs}
\usepackage{multirow}
\usepackage{nicefrac}       % compact symbols for 1/2, etc.
\usepackage{microtype}      % microtypography
\usepackage{xcolor, mdframed}         % colors
\usepackage{amsmath}

\usepackage{orcidlink}
\usepackage{comment}
\usepackage{amssymb}
\usepackage{braket}
\usepackage{xcolor}
\def\BibTeX{{\rm B\kern-.05em{\sc i\kern-.025em b}\kern-.08em
    T\kern-.1667em\lower.7ex\hbox{E}\kern-.125emX}}
\begin{document}

% \title{QFedRansel: A Hybrid Quantum-Enhanced LSTM Framework for Privacy-Preserving Federated Financial Fraud Detection}

\title{ \Huge A Privacy-Preserving Federated Framework with Hybrid Quantum-Enhanced Learning for Financial Fraud Detection}
% \author{\IEEEauthorblockN{Anonymous Authors}}
% \begin{comment}
\author{\IEEEauthorblockN{Abhishek Sawaika\textsuperscript{1}, Swetang Krishna\textsuperscript{2},
Tushar Tomar\textsuperscript{3}, Durga Pritam Suggisetti\textsuperscript{4}, \\
Aditi Lal\textsuperscript{5}, Tanmaya Shrivastav\textsuperscript{3},
Nouhaila Innan\textsuperscript{6,7}, and Muhammad Shafique\textsuperscript{6,7}
\IEEEauthorblockA{
\textsuperscript{1}University of Melbourne, Melbourne, Australia\\
\textsuperscript{2}Trinity College Dublin, Dublin, Ireland\\
\textsuperscript{3}Indian Institute of Technology, Bombay, India\\
\textsuperscript{4}Birla Institute of Technology and Science Pilani, Dubai, UAE\\
\textsuperscript{5}South Asian University, New Delhi, India\\
\textsuperscript{6}eBRAIN Lab, Division of Engineering, New York University Abu Dhabi (NYUAD), Abu Dhabi, UAE\\
\textsuperscript{7}Center for Quantum and Topological Systems (CQTS), NYUAD Research Institute, NYUAD, Abu Dhabi, UAE\\
abhishek.sawaika@student.unimelb.edu.au, krishnsw@tcd.ie,
tomartushar@cse.iitb.ac.in, f20230242@dubai.bits-pilani.ac.in,\\ aditilal@students.sau.ac.in, 21307r016@iitb.ac.in, nouhaila.innan@nyu.edu, muhammad.shafique@nyu.edu\\
}}}
% \end{comment}

\maketitle

\begin{abstract}
Rapid growth of digital transactions has led to a surge in fraudulent activities, challenging traditional detection methods in the financial sector. To tackle this problem, we introduce a specialized federated learning framework that uniquely combines a quantum-enhanced Long Short-Term Memory (LSTM) model with advanced privacy preservation techniques. By integrating quantum layers into the LSTM architecture, our approach adeptly captures complex cross-transactional patterns, resulting in an approximate 5\% performance improvement across key evaluation metrics compared to conventional models. Central to our framework is ``FedRansel'', a novel method designed to defend against poisoning and inference attacks, thereby reducing model degradation and inference accuracy by 4–8\%, compared to standard differential privacy mechanisms. This pseudo-centralized setup with a Quantum LSTM model, enhances fraud detection accuracy and reinforces the security and confidentiality of sensitive financial data.
\end{abstract}


\section{Introduction}
Financial fraud has long posed significant challenges, with documented instances tracing back to antiquity. One of the earliest recorded cases occurred around 300 B.C., when the Greek merchant Hegestratos attempted to commit insurance fraud by sinking his own ship and profiting from the associated loan and cargo resale \cite{investopedia2025history}.
In the modern digital economy, the proliferation of online financial transactions has introduced new vulnerabilities and significantly increased the sophistication of fraudulent activities \cite{reurink2019financial}. As digital systems replace conventional financial infrastructures, adversaries exploit advanced tools to breach security protocols and manipulate transactional data \cite{sun2023digital}. This growing threat landscape underscores the urgent need for fraud detection systems that are not only accurate but also robust and privacy-preserving.

Recent research focusing on Machine Learning (ML) techniques has made considerable progress in addressing this issue. Approaches range from conventional data mining strategies to advanced graph-based models \cite{al2021financial,cheng2024advanced,motie2024financial}, achieving encouraging performance across diverse financial datasets \cite{west2016intelligent,ali2022financial}. However, a key limitation of many such models lies in their dependence on centralized data access. Financial institutions, constrained by stringent data privacy regulations, often find it infeasible to pool sensitive data for centralized training, thereby limiting the applicability of conventional ML pipelines.

Federated Learning (FL) offers a promising alternative to this issue by allowing collaborative model training across decentralized clients without sharing raw data \cite{mcmahan2017communication}. While classical FL solutions \cite{el2024federated,shi2023responsible} provide enhanced privacy guarantees, they often encounter bottlenecks related to computational efficiency, especially when dealing with complex sequential patterns in financial fraud.

To address these challenges, we explore Quantum Computing (QC) as a complementary paradigm. Quantum technologies are increasingly recognized for their potential to tackle high-dimensional, computationally intensive problems in finance, including portfolio optimization, risk assessment, and fraud detection \cite{innan2024financial,innan2024financial2}. By exploiting quantum parallelism and entanglement, quantum-enhanced models can explore complex solution spaces more efficiently than their classical counterparts. According to McKinsey \cite{gschwendtner2025quantum}, QC applications in finance could generate up to $622$ billion in value upon the maturity of fault-tolerant systems.
\begin{table}[h]
    \centering
    \caption{Comparative study of related works.}
    \resizebox{\linewidth}{!}{
    \begin{tabular}{c|c|c|c|c}
    \hline
        \textbf{Related Work} & \textbf{Finance use-case}& \textbf{LSTM} & \textbf{Federated}& \textbf{Quantum}\\
        \hline
        \cite{al2021financial, cheng2024advanced, ali2022financial} & $\checkmark$ &  $\times$  &  $\times$ &  $\times$\\
         \cite{alghofaili2020financial, benchaji2021enhanced}  & $\checkmark$ & $\checkmark$  & $\times$ & $\times$\\
         \cite{innan2024financial, paquet2022quantumleap}  & $\checkmark$ & $\times$ & $\times$ & $\checkmark$\\
         \cite{innan2024qfnn}        & $\checkmark$ & $\times$ & $\checkmark$  &$\checkmark$ \\
        \cite{chen2021federated,chehimi2022quantum, rofougaran2024federated}   & $\times$ &  $\times$ & $\checkmark$ &$\checkmark$\\
        \cite{khan2024quantum, chen2022quantum}   & $\times$ & $\checkmark$ & $\times$ & $\checkmark$\\
         \cite{chehimi2024federated, sen2025qgaphensemble}  & $\times$ & $\checkmark$  & $\checkmark$ & $\checkmark$\\
         \textbf{Ours} & \textbf{$\checkmark$} & \textbf{$\checkmark$} & \textbf{$\checkmark$} & \textbf{$\checkmark$} \\
        \hline
    \end{tabular}
    }
    \label{tab:background}
\end{table}
Recent progress in Quantum Machine Learning (QML) has introduced models such as Variational Quantum Classifiers (VQCs), Quantum Support Vector Classifiers (QSVCs) \cite{cerezo2021variational, li2024quantum,innan2024variational}, capable of learning non-trivial data relationships. Preliminary work in Quantum Federated Learning (QFL) \cite{chen2021federated} has shown impressive accuracy gains on benchmark tasks such as MNIST \cite{li2021quantum}. While some existing studies have explored its application to financial data \cite{innan2024qfnn}, they primarily emphasize the federated learning aspect without integrating dedicated privacy-preserving techniques, leaving a critical gap in secure quantum financial modeling (see Table \ref{tab:background}).

In this work, we propose a novel framework that integrates QC with FL for enhanced and privacy-preserving financial fraud detection. Our main contributions are:\begin{itemize}
    \item \textbf{Quantum-Enhanced LSTM}: We design a quantum-integrated LSTM architecture to study fraud detection in financial data, which enables us to capture complex cross-transactional patterns and thereby improve performance metrics over conventional models.
    \item \textbf{FedRansel Privacy Mechanism}: A novel method, termed FedRansel, is proposed to mitigate privacy and attack issues in FL. We critically analyzed and empirically justified the impact of poisoning and inference attacks on the developed system, with its supremacy over existing differential privacy techniques.
    \item \textbf{Pseudo-Centralized Federated Framework}: With the underlying nature of the developed FedRansel technique, the overall FL effectively functions as a pseudo-centralized setup that ensures robust protection of sensitive data with very minimal overhead on model performance.
\end{itemize}

This document is further organized as follows: In Sec. \ref{sec:2}, we present preliminary concepts and discuss recent related work. Sec. \ref{sec:methodology} details our methodology, model architecture based on Quantum-enhanced Long Short-Term Memory (QLSTM), and comprehensive design analysis. In Sec. \ref{section:fedransel}, we introduce the FedRansel technique and its underlying mathematical formulation. Sec. \ref{sec:exp} describes the experimental settings and procedures employed in our study, while Sec. \ref{sec: res} presents the resulting findings. Finally, closing remarks are provided in Sec. \ref{sec: conc}.

\section{Background} \label{sec:2}
This section provides an overview of key concepts and developments related to our work. We begin by introducing the fundamentals of FL, including its variants and associated privacy threats. We then discuss techniques designed to enhance privacy in FL, followed by recent advances in QML and VQCs. Finally, we review the emerging domain of QFL and its relevance to secure distributed learning.
\subsection{Federated Learning and Its Variants}
\begin{figure}[htpb]
    \centering
\includegraphics[width=1\linewidth]{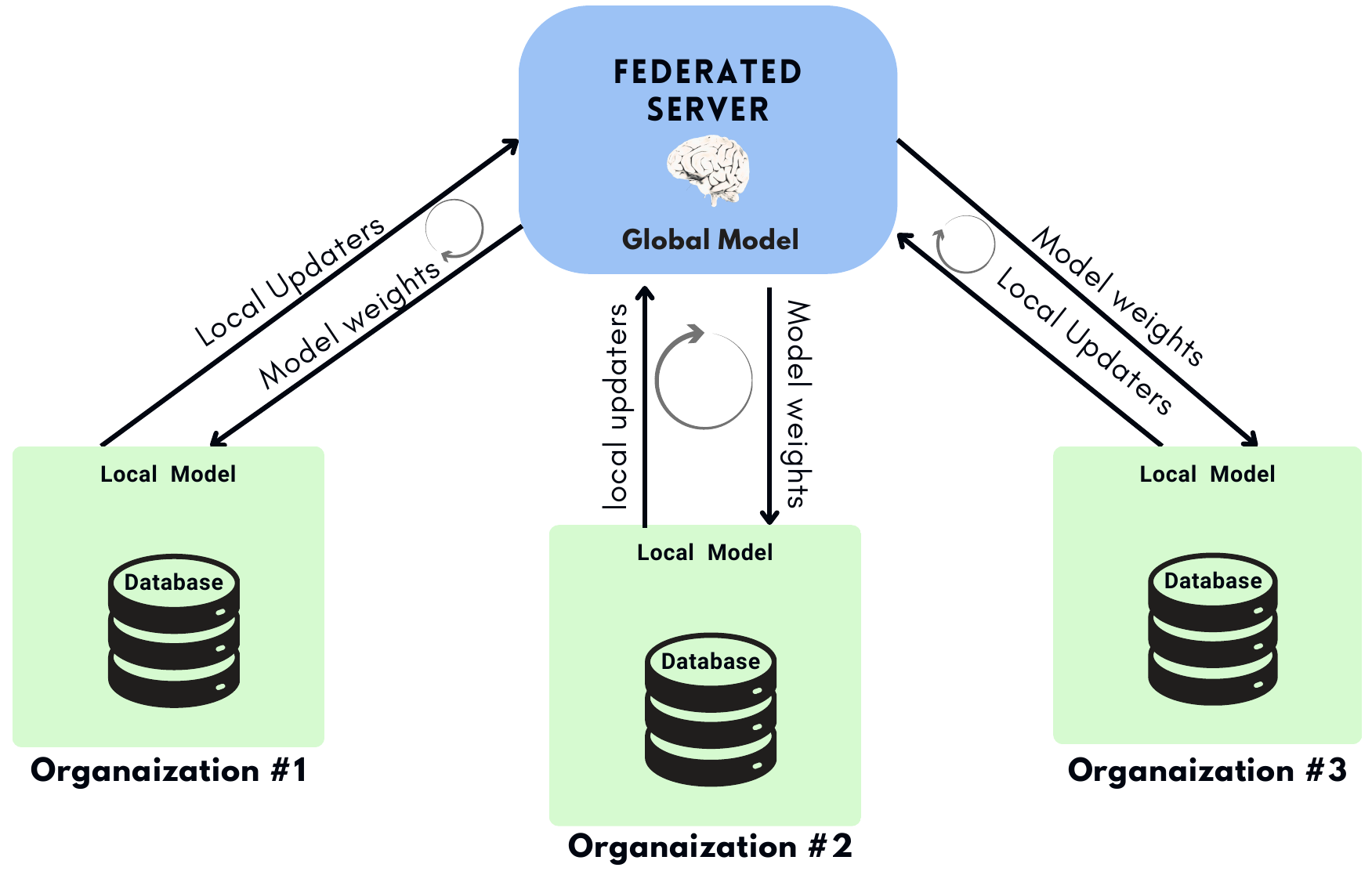}
    \caption{\small Overview of FL, each participating organization trains a local model on private data, while a central federated server aggregates these local updates to construct a collaborative global model without data sharing.}
    \label{fig:fed}
\end{figure}
FL has emerged as a compelling paradigm for collaboratively training machine learning models across distributed data sources, without requiring raw data to be shared with a central server \cite{mcmahan2017communication}. Fig. \ref{fig:fed} shows a simple workflow of FL framework, with three organizations, each having their data stored locally, and a central server for collaborative merging and information transfer between different parties. In every iteration, local models are trained on individual nodes, and learned model parameters are shared with the federated server for merging. Once the information from all nodes is processed, it is sent back for the next round of training. This process continues until the global model converges to the required task.

We can also model this as a fully decentralized peer-to-peer learning \cite{lalitha2018fully}, without the need for a central server. In addition, several variants of FL have been proposed based on different data distributions and coordination requirements, such as:

\begin{itemize} 
\item \textit{Horizontal FL:} is employed when participating entities share a common feature space but possess different user samples \cite{yang2020horizontal}. This is typical when several banks, each with distinct customer bases, collaborate to detect financial fraud, as applicable to our study in this work.
\item \textit{Vertical FL:} applies when participants have common sample IDs but distinct feature sets \cite{liu2024vertical}, such as a collaboration between a bank and a credit scoring agency, each contributing complementary data about the same clients.
\item \textit{Federated Transfer Learning:} addresses situations where both the sample IDs and feature spaces differ, leveraging transfer learning principles to bridge gaps across domains \cite{liu2020secure}.
\end{itemize}

\subsection{Attacks on FL}
\begin{figure*}[htpb]
    \centering
    \includegraphics[width=1\linewidth]{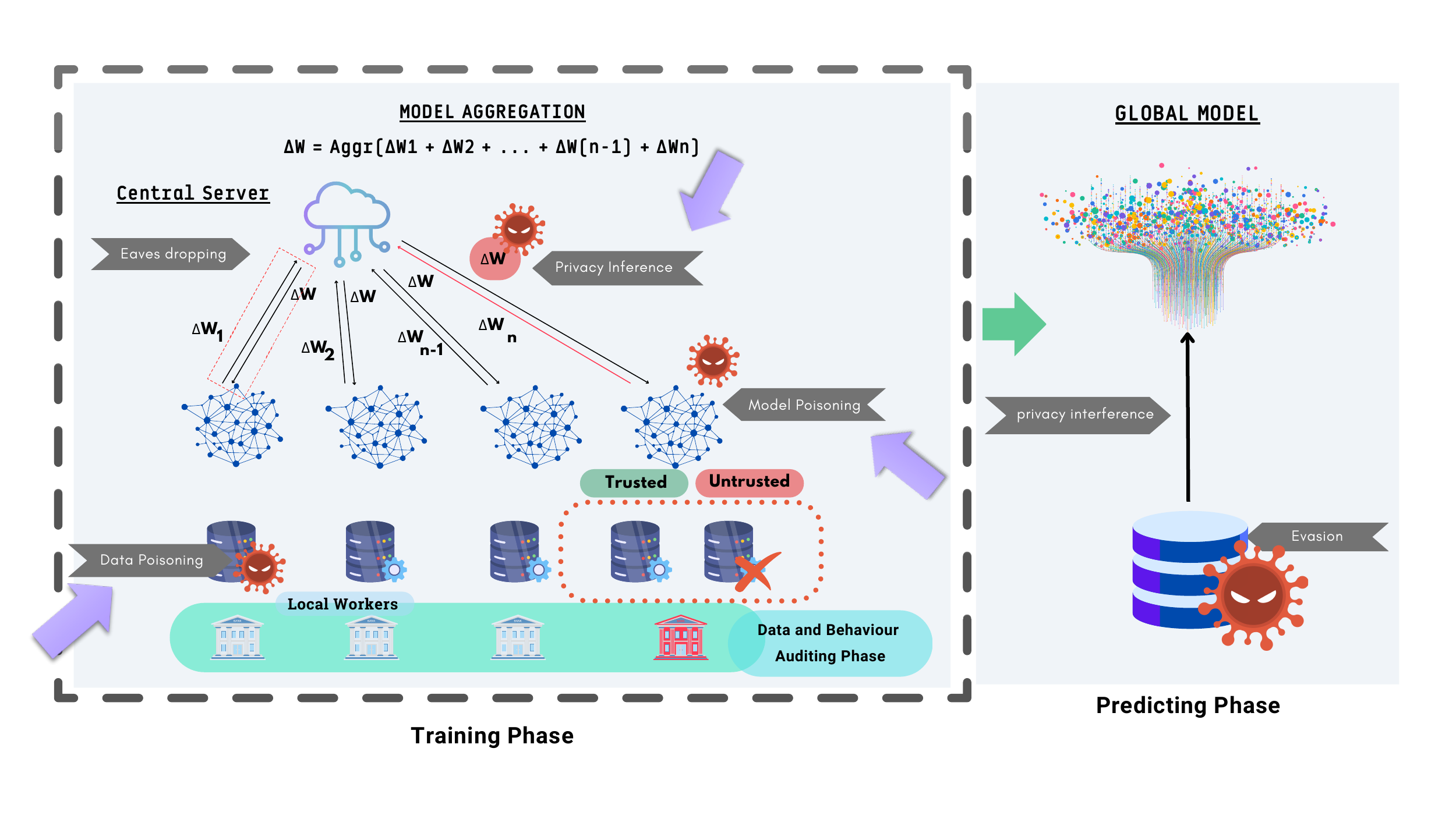}
    \vspace{-1cm}
    \caption{ \small Common attacks and their target realization in FL. The diagram represents a traditional FL setup, including various types of attacks that occur during both the training and prediction phases. It indicates the exact location/step where the attack happens in the overall system. Our focus is on the ones indicated by the arrows during the training phase. $\Delta W's$ here represent the model updates by individual participants.}
    \label{fig:fed-attack}
\end{figure*}
Despite the decentralized nature of FL, several vulnerabilities persist that threaten model integrity and data confidentiality \cite{liu2022threats}. These threats can originate from both the the server and worker sides of the FL architecture (see Fig.~\ref{fig:fed-attack}). In this work, we address two of the most common attacks, namely, poisoning and membership inference attacks. These are described as follows:

\subsubsection{Poisoning Attacks} 
In this attack \cite{xia2023poisoning}, malicious participants can inject adversarial patterns into training data or manipulate model updates to perform targeted attacks and disrupt the performance of the global model. Several defense mechanisms, such as robust aggregation, differential privacy, and statistical filtering, have been proposed; however, these often trade off between the performance and robustness of the model.

\subsubsection{Inference Attacks} 
Inference-based threats \cite{nasr2019comprehensive} commonly include ``property inference'' and ``membership inference'' attacks, where an adversary deduces sensitive attributes of the training data or identifies whether a specific sample was used during training or not, respectively. A standard attack model involves ``model inversion'' \cite{huang2021evaluating}, where adversaries attempt to reconstruct input data by analyzing gradient updates. These attacks present serious concerns in financial applications due to the sensitive nature of transactional data.

\subsection{Privacy-Preserving Techniques in FL}
As seen in previous sections, while FL reduces the need for data centralization, it does not inherently guarantee privacy against all forms of leakage through shared gradients and updates. To strengthen privacy protection, several techniques have been proposed for FL systems, which can be equivalently applied to QFL under classical communication.

\textit{Differential Privacy (DP)} remains the most commonly used approach in FL systems \cite{abadi2016deep, mcmahan2017learning, wei2020federated}. By injecting calibrated noise into gradients or model parameters, DP ensures that individual data records cannot be inferred from the model. In ``central-DP'', noise is added after aggregation on the server, assuming a trusted central entity. Whereas, ``local-DP'' introduces noise at the client level, offering stronger privacy at the expense of utility due to compounded noise levels.

Practical deployments, such as Meta's FL-DP system \cite{stojkovic2022applied}, have revealed that privacy enhancements come at the cost of slower convergence and reduced accuracy. Similarly, recent studies \cite{rofougaran2024federated, li2021quantum} also show that while DP defends effectively against inference threats, it often leads to degraded training efficiency.

Other approaches, such as \textit{Homomorphic Encryption (HE)}, enable computations on encrypted data \cite{acar2018survey}, ensuring that no plaintext data is revealed during training. However, the substantial computational overhead associated with HE presents practical challenges for scalability and latency-sensitive applications. In later sections, we introduce a novel privacy-preserving technique designed to address these challenges and provide a comparative analysis with the widely-used DP approach.

\subsection{QML and Variational Models}
\begin{figure*}[htpb]
    \centering
    \includegraphics[width=\linewidth]{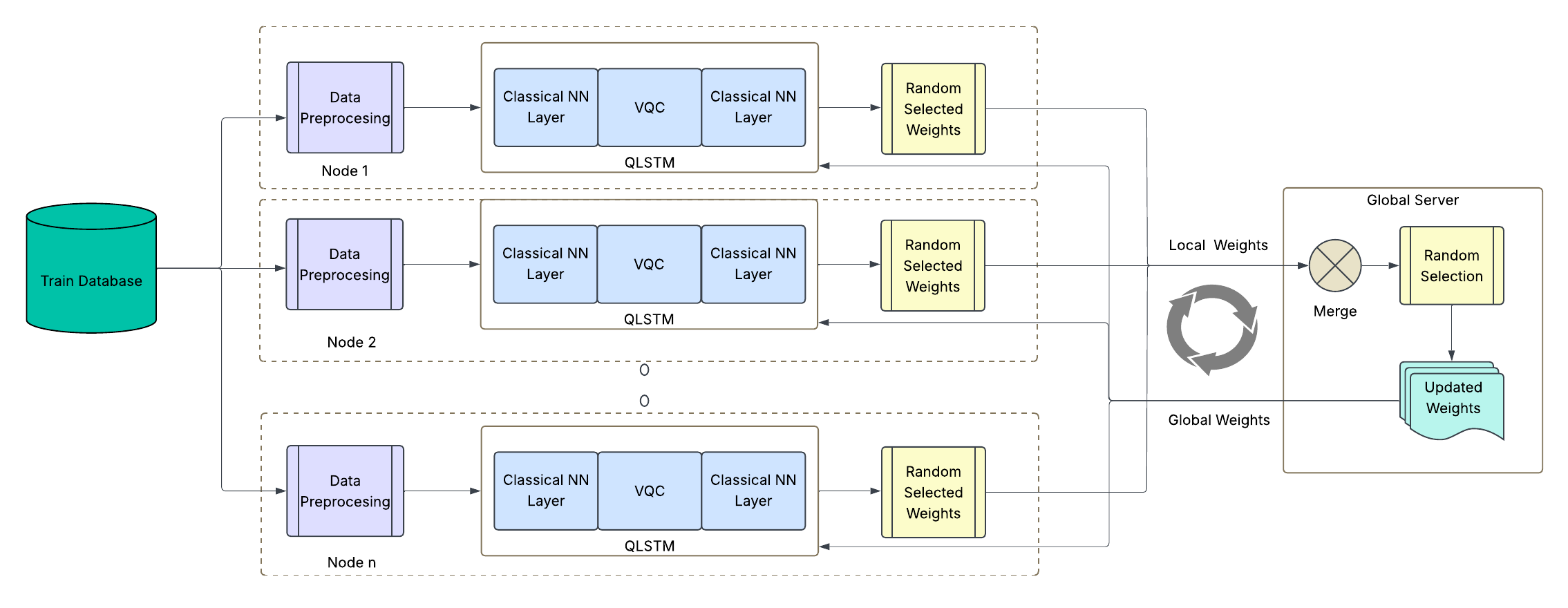}
    \vspace{-0.6cm}
    \caption{\small Training workflow of our framework demonstrating local quantum-enhanced computations, federated aggregation, and secure update exchanges.}
    \label{fig:training}
\end{figure*}

QML combines the computational power of QC with ML algorithms to address problems that are intractable in classical systems. Variational quantum models, particularly VQCs, play a central role in QML by using parameterized quantum gates to encode data and perform optimization tasks in high-dimensional Hilbert spaces.

Models such as QSVMs and quantum kernel methods have shown potential in achieving higher classification accuracy through quantum-enhanced feature representations \cite{rebentrost2014quantum,innan2023enhancing}. These are typically trained using hybrid optimization approaches such as the parameter-shift rule \cite{mitarai2018quantum}.
Recent works have also demonstrated the integration of quantum circuits into classical architectures, such as CNNs and LSTMs, forming hybrid Quantum Neural Networks (QNNs) \cite{khan2024quantum, alrikabi2022face, paquet2022quantumleap,innan2025qnn,innan2025optimizing}. 

These models aim to exploit quantum representations to enhance sequential and structural data modeling. However, they remain challenged by phenomena such as barren plateaus in optimization and hardware-induced noise.
Despite these challenges, the computational advantages of QML market a promising approach for resource-constrained learning tasks that require both expressivity and privacy. Therefore, in this work, we explore the potential power of a quantum-enhanced LSTM architecture for our use case.

\subsection{Quantum Federated Learning}
QFL aims to combine QC with the collaborative, privacy-preserving nature of FL \cite{10651123}. Initial work in this area has focused on proof-of-concept demonstrations, such as the QFL protocols in \cite{chen2021federated}, which applied variational models to classical datasets like CIFAR-10 and Cats and Dogs.

In \cite{li2021quantum}, a decentralized protocol based on blind quantum computing (BQC) was proposed, enabling clients to outsource computations to untrusted quantum servers while maintaining data privacy. Their approach incorporated DP at the gradient level to prevent inversion attacks, demonstrating strong accuracy on MNIST and the Wisconsin Diagnostic Breast Cancer datasets.

More recent work has shifted toward quantum-native communication and processing. For example, recent protocols assume quantum data and communication channels \cite{chehimi2022quantum, wang2024quantum}. 
Some approaches integrate DP into quantum training pipelines, demonstrating resilience against gradient leakage attacks while maintaining model performance \cite{rofougaran2024federated}.

These developments underscore the potential of QFL as a secure and scalable solution for collaborative learning in sensitive domains such as finance and healthcare. However, many of these efforts focus on FL functionality without explicitly integrating robust, privacy-preserving mechanisms, a limitation we address in this work.

\section{Methodology}
\label{sec:methodology}
In this section, we present the design and implementation of the proposed Quantum-LSTM FedRansel model for financial fraud detection within the FL context. Our approach combines state-of-the-art quantum computing techniques with classical LSTM architectures to enhance sequential modeling while preserving data privacy and robustness against adversarial attacks.

\subsection{Design Rationale and Overview}

Recent advances in QML have demonstrated that quantum-enhanced models can capture complex patterns more efficiently than their classical counterparts. Motivated by these results, we integrate VQCs into a classical LSTM framework, yielding a QLSTM network. This is motivated by the one described in \cite{chen2022quantum} with modular changes in layers and architecture being used as per our use case. The overall system is embedded within an FL paradigm where local nodes independently train and then aggregate their models to form a global representation, thereby mitigating risks from inference and poisoning attacks \cite{sen2025qgaphensemble}.

Fig.~\ref{fig:training} illustrates the overall training workflow and highlights the interaction between quantum-enhanced computations and federated aggregation.

\begin{figure}[h]
         \centering
         \includegraphics[width=\linewidth]{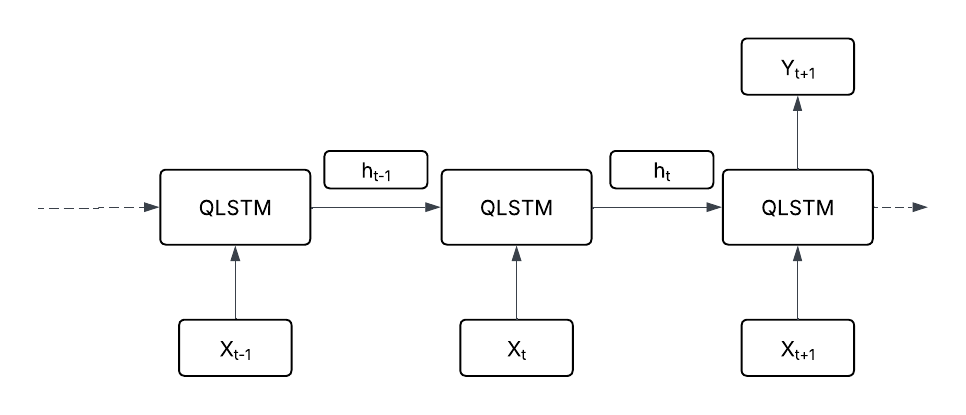}
         \vspace{-0.6cm}
         \caption{ \small QLSTM architecture overview. A three-sequence QLSTM model where each QLSTM cell processes an input at time step $\mathbf{X}_t$, computes a corresponding hidden state $\mathbf{h}_t$, and generates the output $\mathbf{Y}_{t+1}$. This diagram illustrates the sequential stacking of QLSTM cells, facilitating both temporal and federated model integration for enhanced dynamic learning capabilities.}
         \label{fig:arc}
\end{figure}

\subsection{Sequential and Federated Model Integration}

Before being processed by the VQC, each input undergoes dimensional adjustment via linear layers to ensure an optimal representation for quantum operations.
Within the FL setting, each node initializes its own QLSTM model instance—with its own optimizer, loss function, and local dataloaders—and trains on locally available data. The locally trained models are then aggregated across nodes to form a global model. This federated approach leverages diverse data sources while safeguarding privacy and enhancing resilience against vulnerabilities such as inference and poisoning attacks.

The overall QLSTM model is constructed by sequentially stacking QLSTM cells (see Fig.~\ref{fig:arc}), as determined by a hyperparameter specifying the sequence length. This sequential arrangement enables the model to capture long-term dependencies within shuffled financial transactions. For each transaction index $t+1$, the model produces a prediction $Y_{t+1}$ based on the sequence of previous inputs.

\begin{figure}[h]
    \centering
    \includegraphics[width=1\linewidth]{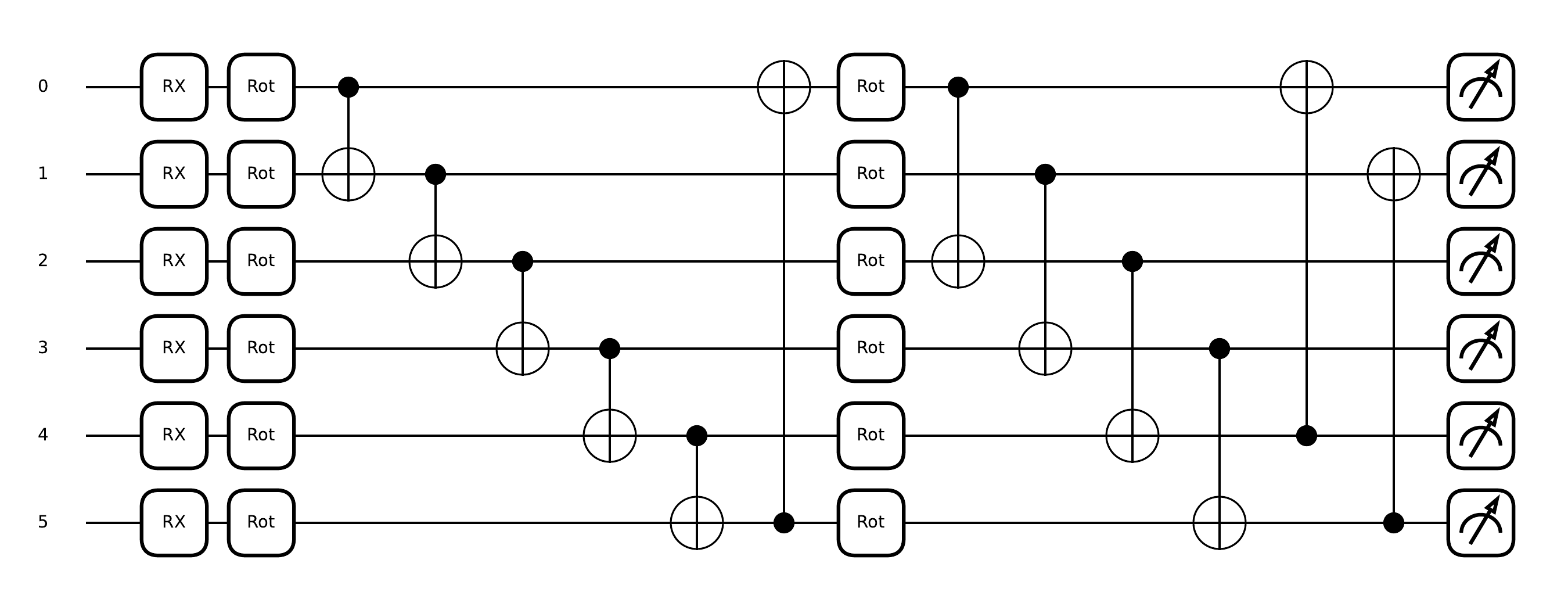}
    \vspace{-0.4cm}
    \caption{ \small The quantum circuit, composed of a sequence of \textbf{RX} operations that implements angle encoding, while the parameterized \textbf{Rot} gate and \textbf{CNOT} operations create a fully entangled variational circuit.}
    \label{fig:vqc}
\end{figure}
\subsection{Quantum-Enhanced LSTM}

The core of our architecture is the QLSTM, which replaces the traditional LSTM gates with VQCs to enable quantum parallelism in processing sequential data.
In classical LSTMs, the cell state is updated through the interaction of four gates: the forget gate, input gate, candidate gate, and output gate, each of which regulates the flow of information through the memory cell. In our QLSTM framework, these gates are redefined via variational quantum circuits. Let $\vec{x}_t \in \mathbb{R}^d$ denote the input vector at time $t$, and $\vec{h}_{t-1}$ denote the hidden state from the previous time step. The quantum gate outputs are computed by a VQC as follows:
\begin{equation} \label{eq:gates}
\{f_t,\, i_t,\, o_t,\, g_t\} = \text{VQC}(\vec{x}_t, \vec{h}_{t-1}),
\end{equation}
where each component is produced by a dedicated quantum circuit block functioning as a gate.

Our VQC design leverages angle encoding followed by layers of entanglement and parameterized rotations. The universal \textbf{Rot} gate is defined as:
\begin{equation}\label{eq:rot}
R(\phi,\theta,\omega) = RZ(\omega)\, RY(\theta)\, RZ(\phi),
\end{equation}
which enables expressive transformations within the Hilbert space. Angle encoding is achieved using \textbf{RX} rotations to embed classical features into quantum states, and subsequent \textbf{CNOT} layers establish full entanglement across qubits (see Fig.~\ref{fig:vqc}).

After obtaining the quantum gate outputs, we update the cell state and hidden state using the standard LSTM formulations:
\begin{align}
c_t &= f_t \odot c_{t-1} + i_t \odot g_t, \label{eq:ct}\\[1mm]
h_t &= o_t \odot \tanh(c_t), \label{eq:ht}
\end{align}
where $\odot$ represents element-wise multiplication, these equations ensure that temporal dependencies are maintained while integrating quantum-computed features.

\begin{figure}[h]
    \centering
    \includegraphics[width=\linewidth]{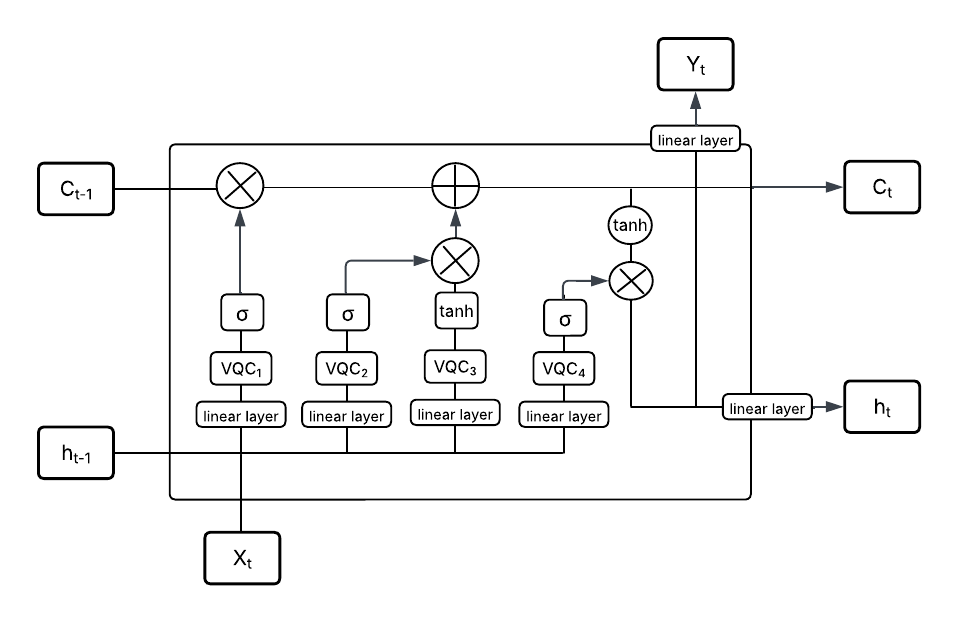}
    \vspace{-0.6cm}
    \caption{\small Architecture of a single QLSTM cell. The cell comprises a classical-to-quantum linear mapping, VQC-based gate operations, and the standard cell state update mechanisms.}
    \label{fig:lstm_cell}
\end{figure}

To illustrate the cell-level integration, Fig.~\ref{fig:lstm_cell} shows the architecture of a single QLSTM cell. In this hybrid design, a classical-to-quantum linear mapping transforms the classical input into a quantum-compatible representation before feeding it into the VQC-based gate operations. This integration enables flexible adaptation to varying qubit requirements and data dimensions while ensuring effective cell state updates according to Equations~\ref{eq:ct} and \ref{eq:ht}.

\section{FedRansel} \label{section:fedransel}

FedRansel introduces a novel FL merging technique tailored for quantum-enhanced privacy and security. It consists of two major steps:
\subsection{Parameter Sampling and Sharing}
During the local training phase at each federated node, only a randomly sampled subset of model parameters is communicated to a centralized global server. This random parameter subset selection significantly mitigates the risk of data reconstruction and inference attacks, maintaining stringent privacy standards.

Let $M_i$ be the set of model parameters at a node $i$ in a collection of N nodes. Then, the shared parameter set $S_i$ is defined based on the local sampling threshold $T_l$ as:
\begin{equation}
    S_i = \{s \ : \ s\in M_i \ and \ |S_i| = \lceil x *|M_i| \rceil\}, \ \text{such that},
    % \ \forall i \in \left[1,N\right]
\end{equation}
\begin{gather*}
   P(X = x) = \frac{1}{(1-T_l)}, \ X\in (0,1] \ \& \ P(s \in S_i) = \frac{1}{|M_i|},
\end{gather*}
\begin{gather*}
  \text{and}  \ if, \ a \in S_i \ \& \ b \in S_i, \ then \ a \neq b.
\end{gather*}

\subsection{Global Parameter Merging}
Upon receiving parameter subsets from multiple nodes, the global server computes an intersection of commonly sampled parameters. Only the averaged common parameters are selected, and a further random subset of these parameters is updated and shared back to the local nodes. This approach effectively safeguards against potential model poisoning and inference attacks while optimizing model learning in a distributed setting. This can be formulated as:
    \begin{equation}
        C = \bigcup_{p \in P} \{\bigcap_{i = 1}^{N} I(p \in S_i) \},
    \end{equation}
    \begin{equation}
        G_a = \{g_q : g_q = \frac{\sum_{i=1}^{N} S_{i}^{q}}{N}, \forall q \in C\},
    \end{equation}
    \begin{equation}
        G_f = \{f_p \ : \ f_p\in G_a \ and \ |G_f| = \lceil T_g *|G_a| \rceil\},
        \end{equation}
        \begin{gather*}
            \& \ P(f_p \in G_f) = 1/|G_a|,
        \end{gather*}
where C is the set of common parameters, P is the set of parameter space, $I(p \in S_i) = \phi$, if false, else $\{p\}$, $G_{a}$ is the set of updated parameter values after global averaging, $S_i^q$ is the value of $q^{th}$ parameter in $S_i$, and $G_f$ is the final set of model parameters sent back to individual nodes after sampling ratio $T_g \in (0,1]$.

The FL process in our approach slightly deviates from existing similar methods \cite{mcmahan2017communication, yuan2024decentralized}. Some of the existing variants based on merging techniques can be found in \cite{qi2024model}. Specifically, our FedRansel method introduces a \textbf{pseudo-centralized} FL framework (see Algorithm \ref{algorithm:alogo}). In this setup, the global server only has limited knowledge of the model parameters, which prevents it from efficiently constructing a global model. However, since only a random subset of the merged global parameters is sent back to the individual nodes, the system does not achieve full decentralization. At any given time, each node may hold a unique set of parameters. It is important to note that our model assumes classical communication between the participating nodes and is not implemented over a purely quantum channel compared to the existing approaches \cite{wang2024quantum}. 

\begin{algorithm}[htpb]
\caption{Pseudo-Centralized Federated Quantum Learning}
\small
\label{algorithm:alogo}
\textbf{Input:} Dataset \(D\), number of clients \(N\), model hyperparameters\\
\textbf{Output:} Trained QLSTM model at local nodes
\BlankLine
% First Procedure: Initialization
\SetKwProg{Proc}{Procedure}{}{end}
\Proc{\(Initialize(D, N, \text{model hyperparameters})\)}{
    \(D_{\text{train}}, D_{\text{test}} \gets \text{Preprocess\_And\_Split}(D)\)\;
    Split \(D_{\text{train}}\) into \(N\) IID subsets for \(N\) clients\;
    \For{each client}{
        Initialize local QLSTM model\;
        Store split data \(\{D_{\text{train}}\}_c\)\;
    }
    \Return Initialized client objects \(\{C\}_N\)\;
}
\BlankLine
% Second Procedure: Training
\SetKwProg{Proc}{Procedure}{}{end}
\Proc{\(FedRansel_{Train}(\{C\}_1^N)\)}{
    \For{each Epoch}{
        \tcp{Local Training}
        \For{each client}{
            Train local QLSTM model\;
            Sample model parameters\;
            Share parameters to global server\;
        }
        \tcp{Global Merge}
        Identify common parameters shared\;
        \If{common parameters exist}{
            Compute average of common parameters\;
            Sample from merged parameters\;
            \tcp{Local Update}
            \For{each client}{
                Update model with newly merged and sampled parameters\;
            }
        }
        \Else{
            Skip global update for this round\;
        }
    }
    \Return Local models\;
}
\end{algorithm}

\section{Data and Experimental Settings} \label{sec:exp}
 This section details the data management strategies, preprocessing techniques, and overall experimental configuration utilized in the study.

\subsection{Data Preprocessing}
Preprocessing procedures are customized based on the specific characteristics of each dataset.

For Dataset 1—the Synthetic Financial Dataset \cite{Dataset1}—dimensionality is reduced using Principal Component Analysis (PCA) with 28 components, which effectively removes multicollinearity while capturing significant variance in a reduced number of dimensions. Randomized shuffling is applied to avoid unintended temporal or positional correlation, thereby preventing any model bias that might arise from an inherent order. Additionally, scaling is performed to ensure compact and efficient representations of transaction sequences.

For Dataset 2—the Bank Fraud Detection Dataset \cite{Dataset2}—class imbalance is addressed through strategic under-sampling techniques that preserve the natural distribution of individual features. Categorical features are transformed via one-hot encoding, and scaling is applied to optimize model training. Due to the limited number of representative features in Dataset 2, PCA is not performed.

\subsection{Optimization and Loss Function}
Model training is performed using classical optimization techniques, such as Adam and Stochastic Gradient Descent (SGD), to navigate the quantum parameter space. The loss function selected is Binary Cross-Entropy with Logits, suitable for binary classification tasks, fraud detection in our case. The Binary Cross-Entropy loss is computed as follows \cite{cox1958regression}:
\begin{equation}
\mathcal{L} = -\frac{1}{m} \sum_{j=1}^{m} \left[ y_j \log(\hat{y}_j) + (1 - y_j) \log(1 - \hat{y}_j) \right],
\end{equation}
\begin{flushleft}
where $y_j$ is the true class label and $\hat{y}_j$ is the predicted probability from the model.
\end{flushleft}
\begin{figure*}[htbp]
     \centering
     \begin{subfigure}{0.5\linewidth}
         \centering
         \includegraphics[width=\textwidth]{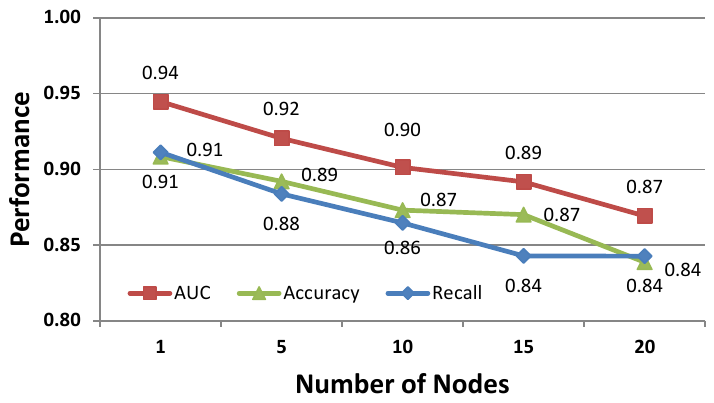}
         \hfill
         \textbf{(a)}
         \label{fig:node1}
     \end{subfigure}%
     \begin{subfigure}{0.5\linewidth}
         \centering
         \includegraphics[width=\textwidth]{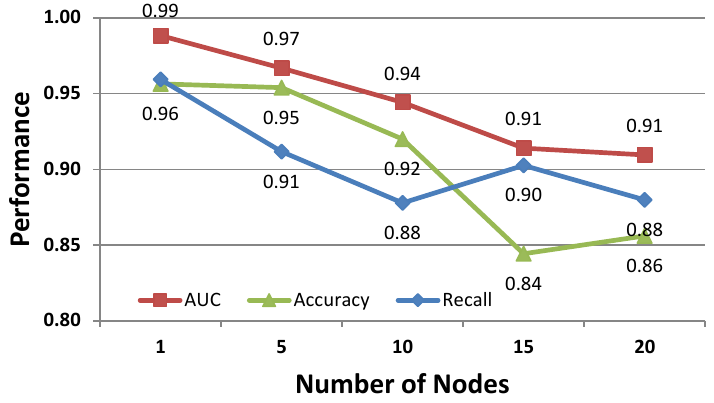}
         \hfill
         \textbf{(b)}
         \label{fig:node2}
     \end{subfigure}
     \vspace{-0.4cm}
    \caption{ \small Federation analysis of the QLSTM model with the FedRansel mechanism, varying the number of participating nodes during the training process. Here, sub-figures (a), (b) represent results for Dataset 1 and 2, respectively. The (number of qubits, depth, sequence length) is set to (9,10,10) for Dataset 1 and (9, 4, 5) for Dataset 2.}
    \label{fig:Nodes}
\end{figure*}
\subsection{Experiment Setup} 

The dataset is partitioned to simulate an FL environment, with Independent and Identically Distributed (IID) sampling as described below.
Let the overall dataset \(D\) be represented as:
\begin{equation}
D = \{ (x_i, y_i) \}_{i=1}^{N}, \quad x_i \in \mathbb{R}^d,\quad y_i \in \{0,1\}.
\end{equation}

Each client \(k \in \mathcal{C}\) receives a local dataset \(D_k \subset D_{\text{train}}\), ensuring:
\begin{equation}
D_{\text{train}} = \bigcup_{k=1}^{K} D_k,\quad D_i \cap D_j = \varnothing \quad \text{for } i \neq j.
\end{equation}

The dataset is split into training and testing subsets:
\begin{equation}
D = D_{\text{train}} \cup D_{\text{test}},\quad D_{\text{train}} \cap D_{\text{test}} = \varnothing.
\end{equation}

Each local dataset \(D_k\) is i.i.d.-sampled from the common underlying distribution \(P_D\). We use \(N = 20\,000\) samples for each dataset. The models are initialized with the respective optimizers, loss functions, and data loaders on each client node.

Experiments are simulated using the default qubit simulator from PennyLane, with local and global sampling thresholds set to \(T_l = 0.8\) and \(T_g = 0.8\), respectively. The QLSTM model is implemented using the Federated Averaging (FedAvg) method for global aggregation~\cite{mcmahan2017communication}.

Table~\ref{tab:params} summarizes the common hyperparameters and experimental setup used across the experiments.

\begin{table}[h]
    \centering
    \caption{ \small Summary of experimental hyperparameters and settings.}
    \label{tab:params}
    \begin{tabular}{c|c}
        \hline
        \textbf{Hyperparameter/Setup} & \textbf{Values (Dataset 1 / Dataset 2)} \\
        \hline
        Learning Rate               & 0.01 / 0.005 \\
        Optimizer                   & Adam \\
        Batch Size                  & 128 / 64 \\
        Dataset Size                & 20K\\
        Train:Test Ratio            & 2:1 \\
        Backend                     & default.qubit Simulator \\
        Hidden State Size           & 10 / 4 \\
        Epochs                      & 50 \\
        Global Rounds               & 5 \\
        Local Sampling Threshold    & 0.8 \\
        Global Sampling Ratio       & 0.8 \\
        \hline
    \end{tabular}
\end{table} 
\section{Results and Discussion} \label{sec: res}

\begin{figure*}[bt]
    \centering
    \includegraphics[width=1\linewidth]{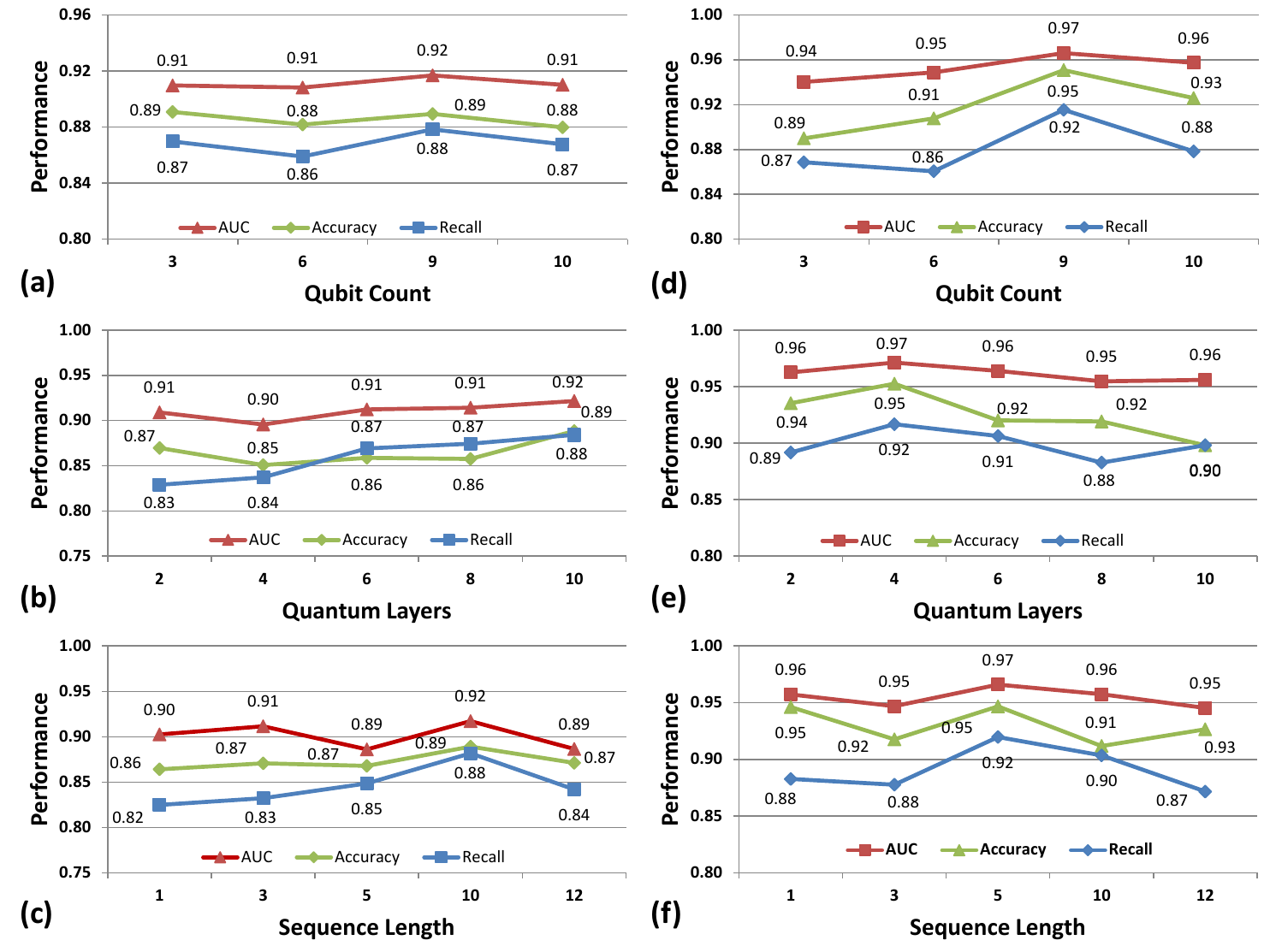}
    \vspace{-0.6cm}
    \caption{ \small Performance trends as a function of varying hyperparameters for \textbf{Dataset 1 (a, b, c)} and \textbf{Dataset 2 (d, e, f)}: In each experiment, one hyperparameter is varied to illustrate its impact on model performance across both datasets, while the other two are fixed at their optimal values, \textbf{(number of qubits, quantum layers, sequence length) = (9, 10, 10)} for Dataset 1 and \textbf{(9, 4, 5)} for Dataset 2.}
    \label{fig:trend}
\end{figure*}
In this section, we provide a detailed analysis of the results obtained from various experiments, emphasizing the potential advantages of our proposed framework over classical models. The model's performance is evaluated using three key metrics: Accuracy, Recall, and AUC scores. Recall measures the model's ability to correctly identify fraudulent transactions, while Accuracy and AUC provide a broader perspective on the model's overall effectiveness. We begin by analyzing the effect of varying the number of participating nodes and then examining the impact of key hyperparameters on model performance. We then proceed with a comparative analysis against existing techniques. Finally, we assess the robustness of the FedRansel mechanism against poisoning and inference attacks.
\subsection{Federated Learning Setup Analysis}
As shown in Fig. \ref{fig:Nodes}, we observe performance degradation as the number of nodes in the FL setup increases. This is primarily due to the smaller dataset available per node for training, which limits the local convergence of each model. To ensure a fair evaluation of the FL environment, we restrict our analysis to experiments with 5 participating nodes, enabling more controlled comparisons across different model configurations.

\subsection{Trend Analysis} \label{subsection:trend}

We examine the performance trends of the model with respect to three key hyperparameters: the number of qubits, the number of quantum layers, and the sequence length. The results for Dataset 1 and Dataset 2 are shown in Fig. \ref{fig:trend}, where each hyperparameter is varied independently, while the other two are held constant at their optimal values based on previous experiments. The trends differ across datasets, but it is crucial to observe how performance evolves as we increase the scalability of our framework. 
\subsubsection{Impact of Number of Qubits}
Increasing the number of qubits enhances the model's ability to handle complex computations, but performance improvements diminish beyond a certain threshold due to factors such as computational limitations and the barren plateau problem. This highlights the scalability of our framework as the number of qubits increases.

For Dataset 1, performance improves with the increasing number of qubits, but only up to 9 qubits. Beyond this point, performance plateaus, indicating that additional qubits do not significantly enhance performance. This suggests that increasing the number of qubits enhances the model's ability to handle more complex computations, but only up to a certain threshold, beyond which further qubits contribute minimally to the improvement.

For Dataset 2, a similar trend is observed: performance increases up to 5 qubits, after which it levels off. Given the smaller feature space in Dataset 2, 5 qubits are sufficient for optimal performance, indicating that the dataset's complexity does not necessitate more qubits. Therefore, while the number of qubits plays a crucial role in enhancing computational capacity, the benefits diminish once a certain threshold is reached, emphasizing the importance of balancing qubit count with dataset complexity.
\subsubsection{Impact of Quantum Layers}
Increasing the number of quantum layers affects the depth of the model, enhancing its expressibility by allowing it to capture more hierarchical features and increasing the potential for overfitting.

For Dataset 1, we observe that performance (particularly AUC and recall) initially dips at a depth of 4 layers but improves as the number of layers increases, reaching a peak at 10 layers. This suggests that the model benefits from increased depth, which enhances its ability to capture complex patterns in Dataset 1. Therefore, the optimal number of quantum layers is 10.

For Dataset 2, the trend is different. Performance improves with increasing quantum layers up to 4 layers, but beyond this point, the model starts to overfit, resulting in a decrease in performance. This indicates that Dataset 2 benefits from fewer layers, with the optimal configuration is 4 layers, allowing the model to capture the necessary features without overfitting.
\subsubsection{Impact of Sequence Length}
The sequence length parameter influences the model's ability to capture correlations across multiple transactions. For Dataset 1, accuracy and recall improve as the sequence length extends up to 10, after which performance begins to degrade. This indicates that longer sequences help capture more complex correlations in Dataset 1, with the best results are achieved at a sequence length of 10.

For Dataset 2, the performance is optimal at a sequence length of 5, which suggests that the smaller feature space in Dataset 2 requires fewer transactions to capture relevant patterns. A performance dip is observed at both ends of sequence length = 5, indicating that excessively short or long sequences degrade performance.

\begin{table*}[htbp] \label{table:performance}
\centering
\caption{ \small Comparative analysis of models and performance metrics evaluated on Dataset1 and Dataset 2.}
\label{tab:performance}
\begin{tabular}{lccc|ccc}
\toprule
\multirow{2}{*}{\textbf{Model}} & \multicolumn{3}{c}{\textbf{Dataset 1}} & \multicolumn{3}{c}{\textbf{Dataset 2}} \\
\cmidrule(lr){2-4} \cmidrule(lr){5-7}
                            & \textbf{Accuracy} & \textbf{Recall} & \textbf{AUC}  & \textbf{Accuracy} & \textbf{Recall} & \textbf{AUC} \\
\midrule
One-Class SVM [Anomaly Detection]         & 0.61     & 0.75   & 0.80 & 0.73     & 0.85   & 0.90 \\
LSTM       & 0.84     & 0.84   & 0.91 & 0.91     & 0.85   & 0.96 \\
QLSTM          & 0.90     & 0.88   & 0.94 & 0.96     & 0.95   & 0.98 \\
QLSTM + FedRansel & 0.89  & 0.88   & 0.92 & 0.95     & 0.92   & 0.97 \\
\bottomrule
\end{tabular}
\end{table*}

\begin{figure*}[htpb]
     \centering
    \begin{subfigure}{0.45\textwidth}
         \centering
   \includegraphics[width=\textwidth]{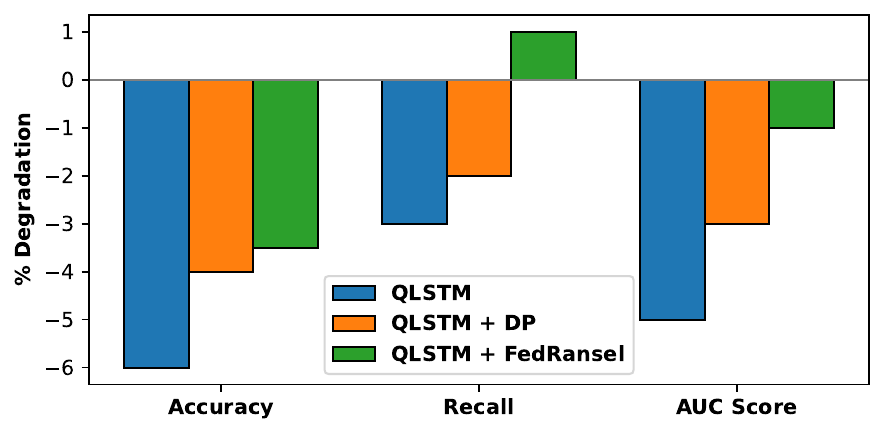}
        % \caption{}
                 \hfill
         \textbf{(a)}
         \label{fig:q1}

     \end{subfigure}%
     \begin{subfigure}{0.45\textwidth}
         \centering
         \includegraphics[width=\textwidth]{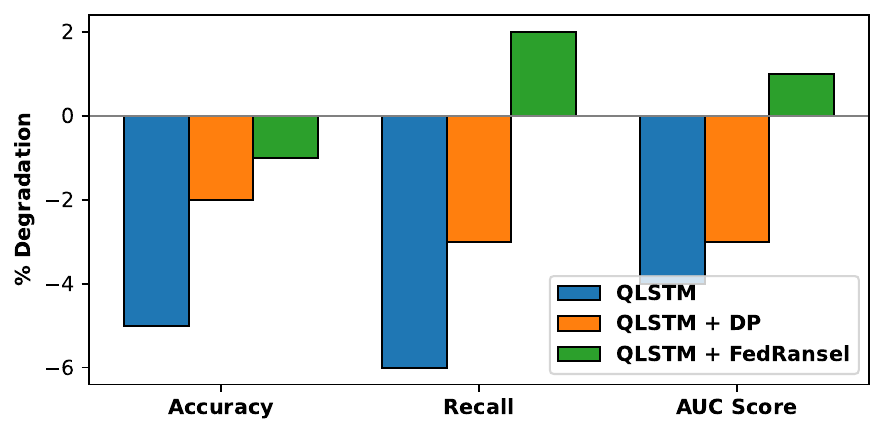}
        %\caption{}
               \hfill
         \textbf{(b)}
         \label{fig:q2}
     \end{subfigure}
     \begin{subfigure}{0.45\textwidth}
         \centering
         \includegraphics[width=\textwidth]{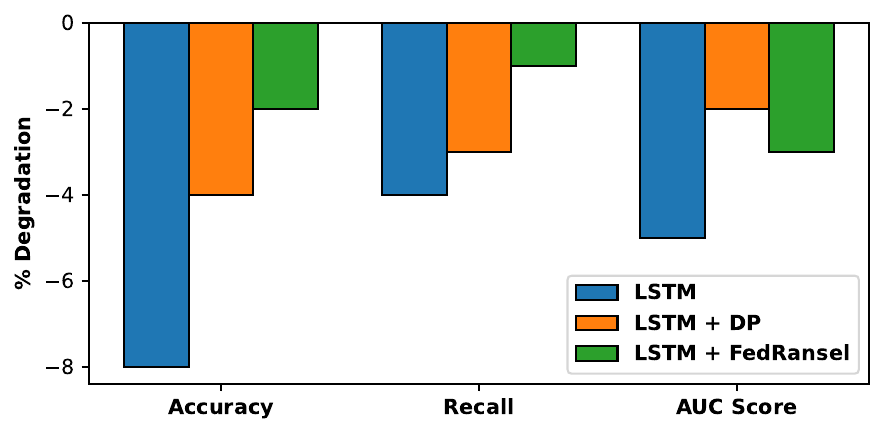}
        %\caption{}
       \hfill
         \textbf{(c)}
        \label{fig:c1}
     \end{subfigure}%
     \begin{subfigure}{0.45\textwidth}
         \centering
         \includegraphics[width=\textwidth]{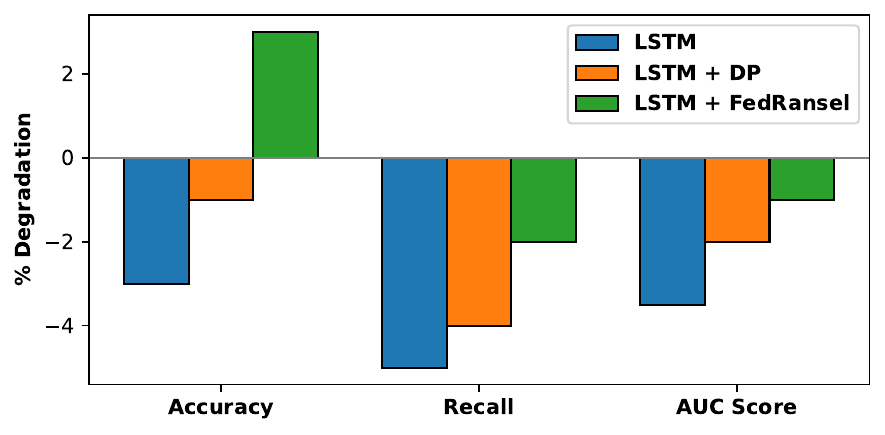}
        %\caption{}
       \hfill
         \textbf{(d)}
    \label{fig:c2}
     \end{subfigure}
\caption{ \small Performance degradation due to poisoning attacks on QLSTM and LSTM models, incorporating DP and our proposed method (FedRansel). The models are evaluated based on three performance metrics: Accuracy, Recall, and AUC Score. Panels (a) and (b) show the results for QLSTM and panels (c) and (d) for classical LSTM, using Datasets 1 and 2, respectively. The degradation is reported as the percentage change in performance, with positive and negative values indicating improvement and degradation, respectively.}   
        \label{fig: Attack}
\end{figure*}
\subsection{Comparative Performance Analysis}

As shown in Table \ref{tab:performance}, we evaluate a comparative analysis of model performance across Dataset 1 and Dataset 2, focusing on three key metrics: Accuracy, Recall, and AUC. We experiment with several models, including a classical LSTM \cite{hochreiter1997long}, which is analogous to the QLSTM model and with a similar number of parameters ($\sim400$), and an anomaly detection technique using One-Class SVM \cite{scholkopf1999support}. The results show that One-Class SVM performs poorly, with accuracy scores of only $0.61$ for Dataset 1 and $0.73$ for Dataset 2.
In contrast, when comparing the QLSTM and LSTM models, we observe substantial performance improvements, particularly with the quantum model. The addition of the FedRansel privacy-preserving module results in a negligible drop of approximately 1\% in performance across both datasets, which is a reasonable tradeoff considering the privacy gains.

The QLSTM outperforms the classical LSTM by approximately 2\% in AUC, 5\% in Accuracy, and 10\% in Recall for Dataset 1, and by 3\% in AUC, 6\% in Accuracy, and 4\% in Recall for Dataset 2. These results demonstrate that, under the given settings, the quantum version of the model slightly outperforms the classical version. While we cannot generalize this conclusion for all scenarios, it is clear that in this specific case, with these particular settings, the QLSTM model shows a performance advantage over the classical LSTM.

\subsection{Privacy Threat Analysis}\label{subsection:attack}

To better understand the impact of attacks on our system, we model and study poisoning and membership inference attacks \cite{xia2023poisoning, bai2024membership}. 

\subsubsection{Attack Model}
For poisoning, we consider both data and model poisoning attacks \cite{fang2020local}, using a static label flip probability of $0.8$ and Poisson noise in model parameters with $\lambda = 0.1$. We simulate membership inference attacks \cite{7958568}. We compare the performance of the FedRansel technique against widely used DP methods at the global server \cite{naseri2020local, sun2019can}, using a norm-bound threshold of $5$ and Gaussian noise with $\delta = 0.2$. 

\subsubsection{%Prevention Analysis
Privacy Preservation Analysis}

Our preliminary results show an approximately 8\% improvement in attack accuracy compared to the DP technique under the membership inference attack.
As shown in Fig. \ref{fig: Attack}-a and b, we observe approximately 6\% degradation in model performance based on accuracy and 5\% based on AUC, for Dataset 1 in the QLSTM model. However, after incorporating the FedRansel mechanism, we reduce this degradation to 2\% and 1\%, respectively, resulting in an overall gain of approximately 4\%, with 2\% coming from DP. Dataset 2 yields slightly better results, showing approximately 6\% less degradation in recall and around 4\% less degradation compared to DP.
We also demonstrate similar trends for the proposed method in the analogous classical model, as shown in Fig. \ref{fig: Attack}-c and d. This further justifies our claims regarding the general robustness and privacy preservation of the FedRansel technique.
In addition to reducing model degradation under attack, our technique also improves performance on specific metrics. This improvement is reflected in the positive values of percentage degradation observed for accuracy, recall, and AUC.

\section{Conclusion} \label{sec: conc}
In this work, we proposed an efficient solution to the problem of fraud detection in financial systems by leveraging carefully engineered data processing techniques and a hybrid quantum-based LSTM architecture. Our empirical study, conducted using quantum simulators, demonstrated a significant ~5\% improvement in performance, compared to a purely classical LSTM model. Additionally, we also demonstrate that our novel privacy-preserving methodology, FedRansel, provides enhanced security against poisoning and membership-inference attacks. Our method generalizes well to both quantum and classical models, with an overall improvement of approximately 4--6\% across various evaluation metrics. Moreover, our empirical evidence suggests that this approach outperforms traditional DP techniques. This work opens doors for further research and application of the proposed technique to other classical and quantum ML systems.
\section*{Acknowledgments}
 This work was supported in part by the NYUAD Center for Quantum and Topological Systems (CQTS), funded by Tamkeen under the NYUAD Research Institute grant CG008, and the Center for Cyber Security (CCS), funded by Tamkeen under the NYUAD Research Institute Award G1104.
\bibliographystyle{IEEEtran}

\bibliography{conference_101719}

% Generated by IEEEtran.bst, version: 1.14 (2015/08/26)
\begin{thebibliography}{10}
\providecommand{\url}[1]{#1}
\csname url@samestyle\endcsname
\providecommand{\newblock}{\relax}
\providecommand{\bibinfo}[2]{#2}
\providecommand{\BIBentrySTDinterwordspacing}{\spaceskip=0pt\relax}
\providecommand{\BIBentryALTinterwordstretchfactor}{4}
\providecommand{\BIBentryALTinterwordspacing}{\spaceskip=\fontdimen2\font plus
\BIBentryALTinterwordstretchfactor\fontdimen3\font minus \fontdimen4\font\relax}
\providecommand{\BIBforeignlanguage}[2]{{%
\expandafter\ifx\csname l@#1\endcsname\relax
\typeout{** WARNING: IEEEtran.bst: No hyphenation pattern has been}%
\typeout{** loaded for the language `#1'. Using the pattern for}%
\typeout{** the default language instead.}%
\else
\language=\csname l@#1\endcsname
\fi
#2}}
\providecommand{\BIBdecl}{\relax}
\BIBdecl

\bibitem{investopedia2025history}
I.~Staff, ``A history of financial fraud: America's first financial fraudsters,'' \emph{Investopedia}, 2025.

\bibitem{reurink2019financial}
A.~Reurink, ``Financial fraud: A literature review,'' \emph{Contemporary topics in finance: A collection of literature surveys}, pp. 79--115, 2019.

\bibitem{sun2023digital}
G.~Sun, T.~Li, Y.~Ai, and Q.~Li, ``Digital finance and corporate financial fraud,'' \emph{International Review of Financial Analysis}, vol.~87, p. 102566, 2023.

\bibitem{al2021financial}
K.~G. Al-Hashedi and P.~Magalingam, ``Financial fraud detection applying data mining techniques: A comprehensive review from 2009 to 2019,'' \emph{Computer Science Review}, vol.~40, p. 100402, 2021.

\bibitem{cheng2024advanced}
Y.~Cheng, J.~Guo, S.~Long, Y.~Wu, M.~Sun, and R.~Zhang, ``Advanced financial fraud detection using gnn-cl model,'' in \emph{2024 International Conference on Computers, Information Processing and Advanced Education (CIPAE)}.\hskip 1em plus 0.5em minus 0.4em\relax IEEE, 2024, pp. 453--460.

\bibitem{motie2024financial}
S.~Motie and B.~Raahemi, ``Financial fraud detection using graph neural networks: A systematic review,'' \emph{Expert Systems with Applications}, vol. 240, p. 122156, 2024.

\bibitem{west2016intelligent}
J.~West and M.~Bhattacharya, ``Intelligent financial fraud detection: a comprehensive review,'' \emph{Computers \& security}, vol.~57, pp. 47--66, 2016.

\bibitem{ali2022financial}
A.~Ali, S.~Abd~Razak, S.~H. Othman, T.~A.~E. Eisa, A.~Al-Dhaqm, M.~Nasser, T.~Elhassan, H.~Elshafie, and A.~Saif, ``Financial fraud detection based on machine learning: a systematic literature review,'' \emph{Applied Sciences}, vol.~12, no.~19, p. 9637, 2022.

\bibitem{mcmahan2017communication}
B.~McMahan, E.~Moore, D.~Ramage, S.~Hampson, and B.~A. y~Arcas, ``Communication-efficient learning of deep networks from decentralized data,'' in \emph{Artificial intelligence and statistics}.\hskip 1em plus 0.5em minus 0.4em\relax PMLR, 2017, pp. 1273--1282.

\bibitem{el2024federated}
T.~El~Hallal and Y.~El~Mourabit, ``Federated learning for credit card fraud detection: Key fundamentals and emerging trends,'' in \emph{2024 International Conference on Circuit, Systems and Communication (ICCSC)}.\hskip 1em plus 0.5em minus 0.4em\relax IEEE, 2024, pp. 1--6.

\bibitem{shi2023responsible}
Y.~Shi, H.~Song, and J.~Xu, ``Responsible and effective federated learning in financial services: A comprehensive survey,'' in \emph{2023 62nd IEEE Conference on Decision and Control (CDC)}.\hskip 1em plus 0.5em minus 0.4em\relax IEEE, 2023, pp. 4229--4236.

\bibitem{innan2024financial}
N.~Innan, A.~Sawaika, A.~Dhor, S.~Dutta, S.~Thota, H.~Gokal, N.~Patel, M.~A.-Z. Khan, I.~Theodonis, and M.~Bennai, ``Financial fraud detection using quantum graph neural networks,'' \emph{Quantum Machine Intelligence}, vol.~6, no.~1, p.~7, 2024.

\bibitem{innan2024financial2}
N.~Innan, M.~A.-Z. Khan, and M.~Bennai, ``Financial fraud detection: a comparative study of quantum machine learning models,'' \emph{International Journal of Quantum Information}, vol.~22, no.~02, p. 2350044, 2024.

\bibitem{gschwendtner2025quantum}
M.~Gschwendtner, N.~Morgan, and H.~Soller, ``Quantum technology use cases as fuel for value in finance,'' \emph{McKinsey Digital}, 2025.

\bibitem{alghofaili2020financial}
Y.~Alghofaili, A.~Albattah, and M.~A. Rassam, ``A financial fraud detection model based on lstm deep learning technique,'' \emph{Journal of Applied Security Research}, vol.~15, no.~4, pp. 498--516, 2020.

\bibitem{benchaji2021enhanced}
I.~Benchaji, S.~Douzi, B.~El~Ouahidi, and J.~Jaafari, ``Enhanced credit card fraud detection based on attention mechanism and lstm deep model,'' \emph{Journal of Big Data}, vol.~8, pp. 1--21, 2021.

\bibitem{paquet2022quantumleap}
E.~Paquet and F.~Soleymani, ``Quantumleap: Hybrid quantum neural network for financial predictions,'' \emph{Expert Systems with Applications}, vol. 195, p. 116583, 2022.

\bibitem{innan2024qfnn}
N.~Innan, A.~Marchisio, M.~Shafique, and M.~Bennai, ``Qfnn-ffd: Quantum federated neural network for financial fraud detection,'' \emph{arXiv preprint arXiv:2404.02595}, 2024.

\bibitem{chen2021federated}
S.~Y.-C. Chen and S.~Yoo, ``Federated quantum machine learning,'' \emph{Entropy}, vol.~23, no.~4, p. 460, 2021.

\bibitem{chehimi2022quantum}
M.~Chehimi and W.~Saad, ``Quantum federated learning with quantum data,'' in \emph{ICASSP 2022-2022 IEEE International Conference on Acoustics, Speech and Signal Processing (ICASSP)}.\hskip 1em plus 0.5em minus 0.4em\relax IEEE, 2022, pp. 8617--8621.

\bibitem{rofougaran2024federated}
R.~Rofougaran, S.~Yoo, H.-H. Tseng, and S.~Y.-C. Chen, ``Federated quantum machine learning with differential privacy,'' in \emph{ICASSP 2024-2024 IEEE International Conference on Acoustics, Speech and Signal Processing (ICASSP)}.\hskip 1em plus 0.5em minus 0.4em\relax IEEE, 2024, pp. 9811--9815.

\bibitem{khan2024quantum}
S.~Z. Khan, N.~Muzammil, S.~Ghafoor, H.~Khan, S.~M.~H. Zaidi, A.~J. Aljohani, and I.~Aziz, ``Quantum long short-term memory (qlstm) vs. classical lstm in time series forecasting: a comparative study in solar power forecasting,'' \emph{Frontiers in Physics}, vol.~12, p. 1439180, 2024.

\bibitem{chen2022quantum}
S.~Y.-C. Chen, S.~Yoo, and Y.-L.~L. Fang, ``Quantum long short-term memory,'' in \emph{Icassp 2022-2022 IEEE international conference on acoustics, speech and signal processing (ICASSP)}.\hskip 1em plus 0.5em minus 0.4em\relax IEEE, 2022, pp. 8622--8626.

\bibitem{chehimi2024federated}
M.~Chehimi, S.~Y.-C. Chen, W.~Saad, and S.~Yoo, ``Federated quantum long short-term memory (fedqlstm),'' \emph{Quantum Machine Intelligence}, vol.~6, no.~2, p.~43, 2024.

\bibitem{sen2025qgaphensemble}
A.~Sen, U.~Sen, M.~Paul, A.~P. Padhy, S.~Sai, A.~Mallik, and C.~Mallick, ``Qgaphensemble: Combining hybrid qlstm network ensemble via adaptive weighting for short term weather forecasting,'' \emph{arXiv preprint arXiv:2501.10866}, 2025.

\bibitem{cerezo2021variational}
M.~Cerezo, A.~Arrasmith, R.~Babbush, S.~C. Benjamin, S.~Endo, K.~Fujii, J.~R. McClean, K.~Mitarai, X.~Yuan, L.~Cincio \emph{et~al.}, ``Variational quantum algorithms,'' \emph{Nature Reviews Physics}, vol.~3, no.~9, pp. 625--644, 2021.

\bibitem{li2024quantum}
J.~Li, Y.~Li, J.~Song, J.~Zhang, and S.~Zhang, ``Quantum support vector machine for classifying noisy data,'' \emph{IEEE Transactions on Computers}, 2024.

\bibitem{innan2024variational}
N.~Innan and M.~Bennai, ``A variational quantum perceptron with grover’s algorithm for efficient classification,'' \emph{Physica Scripta}, vol.~99, no.~5, p. 055120, 2024.

\bibitem{li2021quantum}
W.~Li, S.~Lu, and D.-L. Deng, ``Quantum federated learning through blind quantum computing,'' \emph{Science China Physics, Mechanics \& Astronomy}, vol.~64, no.~10, p. 100312, 2021.

\bibitem{lalitha2018fully}
A.~Lalitha, S.~Shekhar, T.~Javidi, and F.~Koushanfar, ``Fully decentralized federated learning,'' in \emph{Third workshop on bayesian deep learning (NeurIPS)}, vol.~12, 2018.

\bibitem{yang2020horizontal}
Q.~Yang, Y.~Liu, Y.~Cheng, Y.~Kang, T.~Chen, and H.~Yu, ``Horizontal federated learning,'' in \emph{Federated learning}.\hskip 1em plus 0.5em minus 0.4em\relax Springer, 2020, pp. 49--67.

\bibitem{liu2024vertical}
Y.~Liu, Y.~Kang, T.~Zou, Y.~Pu, Y.~He, X.~Ye, Y.~Ouyang, Y.-Q. Zhang, and Q.~Yang, ``Vertical federated learning: Concepts, advances, and challenges,'' \emph{IEEE Transactions on Knowledge and Data Engineering}, vol.~36, no.~7, pp. 3615--3634, 2024.

\bibitem{liu2020secure}
Y.~Liu, Y.~Kang, C.~Xing, T.~Chen, and Q.~Yang, ``A secure federated transfer learning framework,'' \emph{IEEE Intelligent Systems}, vol.~35, no.~4, pp. 70--82, 2020.

\bibitem{liu2022threats}
P.~Liu, X.~Xu, and W.~Wang, ``Threats, attacks and defenses to federated learning: issues, taxonomy and perspectives,'' \emph{Cybersecurity}, vol.~5, no.~1, p.~4, 2022.

\bibitem{xia2023poisoning}
G.~Xia, J.~Chen, C.~Yu, and J.~Ma, ``Poisoning attacks in federated learning: A survey,'' \emph{Ieee Access}, vol.~11, pp. 10\,708--10\,722, 2023.

\bibitem{nasr2019comprehensive}
M.~Nasr, R.~Shokri, and A.~Houmansadr, ``Comprehensive privacy analysis of deep learning: Passive and active white-box inference attacks against centralized and federated learning,'' in \emph{2019 IEEE symposium on security and privacy (SP)}.\hskip 1em plus 0.5em minus 0.4em\relax IEEE, 2019, pp. 739--753.

\bibitem{huang2021evaluating}
Y.~Huang, S.~Gupta, Z.~Song, K.~Li, and S.~Arora, ``Evaluating gradient inversion attacks and defenses in federated learning,'' \emph{Advances in neural information processing systems}, vol.~34, pp. 7232--7241, 2021.

\bibitem{abadi2016deep}
M.~Abadi, A.~Chu, I.~Goodfellow, H.~B. McMahan, I.~Mironov, K.~Talwar, and L.~Zhang, ``Deep learning with differential privacy,'' in \emph{Proceedings of the 2016 ACM SIGSAC conference on computer and communications security}, 2016, pp. 308--318.

\bibitem{mcmahan2017learning}
H.~B. McMahan, D.~Ramage, K.~Talwar, and L.~Zhang, ``Learning differentially private recurrent language models,'' \emph{arXiv preprint arXiv:1710.06963}, 2017.

\bibitem{wei2020federated}
K.~Wei, J.~Li, M.~Ding, C.~Ma, H.~H. Yang, F.~Farokhi, S.~Jin, T.~Q. Quek, and H.~V. Poor, ``Federated learning with differential privacy: Algorithms and performance analysis,'' \emph{IEEE transactions on information forensics and security}, vol.~15, pp. 3454--3469, 2020.

\bibitem{stojkovic2022applied}
B.~Stojkovic, J.~Woodbridge, Z.~Fang, J.~Cai, A.~Petrov, S.~Iyer, D.~Huang, P.~Yau, A.~S. Kumar, H.~Jawa \emph{et~al.}, ``Applied federated learning: Architectural design for robust and efficient learning in privacy aware settings,'' \emph{arXiv preprint arXiv:2206.00807}, 2022.

\bibitem{acar2018survey}
A.~Acar, H.~Aksu, A.~S. Uluagac, and M.~Conti, ``A survey on homomorphic encryption schemes: Theory and implementation,'' \emph{ACM Computing Surveys (Csur)}, vol.~51, no.~4, pp. 1--35, 2018.

\bibitem{rebentrost2014quantum}
P.~Rebentrost, M.~Mohseni, and S.~Lloyd, ``Quantum support vector machine for big data classification,'' \emph{Physical review letters}, vol. 113, no.~13, p. 130503, 2014.

\bibitem{innan2023enhancing}
N.~Innan, M.~A.-Z. Khan, B.~Panda, and M.~Bennai, ``Enhancing quantum support vector machines through variational kernel training,'' \emph{Quantum Information Processing}, vol.~22, no.~10, p. 374, 2023.

\bibitem{mitarai2018quantum}
K.~Mitarai, M.~Negoro, M.~Kitagawa, and K.~Fujii, ``Quantum circuit learning,'' \emph{Physical Review A}, vol.~98, no.~3, p. 032309, 2018.

\bibitem{alrikabi2022face}
H.~ALRikabi, I.~A. Aljazaery, J.~S. Qateef, A.~H.~M. Alaidi, and M.~Roa’a, ``Face patterns analysis and recognition system based on quantum neural network qnn,'' \emph{iJIM}, vol.~16, no.~08, p.~35, 2022.

\bibitem{innan2025qnn}
N.~Innan, B.~K. Behera, S.~Al-Kuwari, and A.~Farouk, ``Qnn-vrcs: A quantum neural network for vehicle road cooperation systems,'' \emph{IEEE Transactions on Intelligent Transportation Systems}, 2025.

\bibitem{innan2025optimizing}
K.~Dave, N.~Innan, B.~K. Behera, S.~Mumtaz, S.~Al-Kuwari, A.~Farouk \emph{et~al.}, ``Optimizing low-energy carbon iiot systems with quantum algorithms: Performance evaluation and noise robustness,'' \emph{IEEE Internet of Things Journal}, 2025.

\bibitem{10651123}
N.~Innan, M.~A.-Z. Khan, A.~Marchisio, M.~Shafique, and M.~Bennai, ``Fedqnn: Federated learning using quantum neural networks,'' in \emph{2024 International Joint Conference on Neural Networks (IJCNN)}, 2024, pp. 1--9.

\bibitem{wang2024quantum}
T.~Wang, H.-H. Tseng, and S.~Yoo, ``Quantum federated learning with quantum networks,'' in \emph{ICASSP 2024-2024 IEEE International Conference on Acoustics, Speech and Signal Processing (ICASSP)}.\hskip 1em plus 0.5em minus 0.4em\relax IEEE, 2024, pp. 13\,401--13\,405.

\bibitem{yuan2024decentralized}
L.~Yuan, Z.~Wang, L.~Sun, P.~S. Yu, and C.~G. Brinton, ``Decentralized federated learning: A survey and perspective,'' \emph{IEEE Internet of Things Journal}, 2024.

\bibitem{qi2024model}
P.~Qi, D.~Chiaro, A.~Guzzo, M.~Ianni, G.~Fortino, and F.~Piccialli, ``Model aggregation techniques in federated learning: A comprehensive survey,'' \emph{Future Generation Computer Systems}, vol. 150, pp. 272--293, 2024.

\bibitem{Dataset1}
\BIBentryALTinterwordspacing
\emph{Fraud detection bank dataset 20K records binary}. [Online]. Available: \url{https://www.kaggle.com/datasets/volodymyrgavrysh/fraud-detection-bank-dataset-20k-records-binary}
\BIBentrySTDinterwordspacing

\bibitem{Dataset2}
\BIBentryALTinterwordspacing
\emph{Synthetic Financial Datasets For Fraud Detection}. [Online]. Available: \url{https://www.kaggle.com/datasets/ealaxi/paysim1}
\BIBentrySTDinterwordspacing

\bibitem{cox1958regression}
D.~R. Cox, ``The regression analysis of binary sequences,'' \emph{Journal of the Royal Statistical Society Series B: Statistical Methodology}, vol.~20, no.~2, pp. 215--232, 1958.

\bibitem{hochreiter1997long}
S.~Hochreiter and J.~Schmidhuber, ``Long short-term memory,'' \emph{Neural computation}, vol.~9, no.~8, pp. 1735--1780, 1997.

\bibitem{scholkopf1999support}
B.~Sch{\"o}lkopf, R.~C. Williamson, A.~Smola, J.~Shawe-Taylor, and J.~Platt, ``Support vector method for novelty detection,'' \emph{Advances in neural information processing systems}, vol.~12, 1999.

\bibitem{bai2024membership}
L.~Bai, H.~Hu, Q.~Ye, H.~Li, L.~Wang, and J.~Xu, ``Membership inference attacks and defenses in federated learning: A survey,'' \emph{ACM Computing Surveys}, vol.~57, no.~4, pp. 1--35, 2024.

\bibitem{fang2020local}
M.~Fang, X.~Cao, J.~Jia, and N.~Gong, ``Local model poisoning attacks to $\{$Byzantine-Robust$\}$ federated learning,'' in \emph{29th USENIX security symposium (USENIX Security 20)}, 2020, pp. 1605--1622.

\bibitem{7958568}
R.~Shokri, M.~Stronati, C.~Song, and V.~Shmatikov, ``Membership inference attacks against machine learning models,'' in \emph{2017 IEEE Symposium on Security and Privacy (SP)}, 2017, pp. 3--18.

\bibitem{naseri2020local}
M.~Naseri, J.~Hayes, and E.~De~Cristofaro, ``Local and central differential privacy for robustness and privacy in federated learning,'' \emph{arXiv preprint arXiv:2009.03561}, 2020.

\bibitem{sun2019can}
Z.~Sun, P.~Kairouz, A.~T. Suresh, and H.~B. McMahan, ``Can you really backdoor federated learning?'' \emph{arXiv preprint arXiv:1911.07963}, 2019.

\end{thebibliography}

\end{document}